\documentclass[12pt]{iopart}
\begin{document}

\title[Phase theory for confined 1D fermions]{Phase theory and critical 
exponents for the Tomonaga-Luttinger model with harmonic confinement}

\date{\today}

\author{Gao Xianlong and W. Wonneberger}

\address{Abteilung f\"ur Mathematische Physik, Universit\"at Ulm, D89069 Ulm, Germany}

\begin{abstract}
A phase operator formulation for a recent model of interacting one-dimensional fermions in a 
harmonic trap is developed. The resulting theory is similar to the corresponding approach for
the Luttinger model with open boundary conditions (OBC). However, in place of the spatial 
coordinate $z$, 
a dimensionless variable $u$ defined on the unit circle appears as argument of the phase fields
and $u$ is non-linearly related to $z$. Furthermore, form factors appear which reflect the 
harmonic trap geometry. The theory is applied to calculate one-particle correlation functions. 
In a 
properly defined thermodynamic limit, bulk and boundary critical exponents are calculated 
for the static two-point correlation function and the dynamic local correlation function. The
local spectral density is also considered. The critical exponents found are in agreement 
with those known for OBC with the exception of the boundary scaling exponent $\Delta _\perp$.
\end{abstract}


\eads{\mailto{xianlong.gao@physik.uni-ulm.de},
\mailto{wolfgang.wonneberger@physik.uni-ulm.de}}\maketitle

\section{Introduction}

The concepts underlying the Luttinger model (LM) \cite{T50,L63,ML65,Hal81} (for reviews cf. 
\cite{E79,So79,
V95,Sch95}) find increasing application in the theory of quasi one-dimensional ultracold
quantum gases. Using the harmonic fluid model of \cite{Hal81a}, properties  of a one-dimensional
trapped Bose gas were investigated in \cite{MLE98}. Related concepts were also used in the 
theory of
interacting bosons in one-dimensional optical lattices \cite{BBZ03,C03}. In \cite{RZ03}, 
mass-composition separation as analogue of spin-charge separation in the LM
was studied for a two-component neutral Fermi gas in a one-dimensional harmonic trap (cf. also
\cite{GW2}). A 
detailed description of the harmonic fluid approach to one-dimensional quantum gases is given 
in \cite{Caz04}. 

The present authors formulated a theory of interacting one-dimensional fermions in a harmonic 
atom
trap \cite{GW2,WW} based on bosonization in analogy to the LM. The theory was initiated by the 
prospect to create experimentally a highly 
degenerate Fermi gas of neutral atoms in a quasi one-dimensional harmonic trap. A possible route
to achieve this aim is microtrap technology (cf.\cite{VFP98,FGZ98,DCS99,RHH99,Ott01}).
This will allow to study non-Fermi liquid behavior of interacting one-dimensional Fermi 
gases without the complications due to contacts and impurities. 

The model in \cite{GW2,WW}, which can be termed Tomonaga-Luttinger model with harmonic confinement,
lacks so far a phase operator formulation known for many models of one-dimensional interacting 
fermions. Probably best known for such a phase operator formulation is the Tomonaga-Luttinger 
model with periodic boundary conditions. A phase formulation for the corresponding
model with open boundary conditions (OBC), which belongs to another universality class 
\cite{VYG00}, is found in \cite{FG95} and in particular in \cite{MEJ97}. Related phase theories 
also exist for the case of interacting one-dimensional bosons \cite{Hal81a,Caz04,Caz02}.

In general, a phase operator approach to one-dimensional fermions comprises several features:

\begin{itemize}
\item[i)]
Extending the linearized dispersion near the Fermi energy with respect to fermionic creation 
and annihilation operators $\hat{c}^+ _{k \sigma}$ and $\hat{c}_{k \sigma}$ to all values of 
$k$ and adding an anomalous vacuum of filled negative 
energy states, Kronig's identity \cite{deK35} can be used to transform the free 
Hamiltonian into a free bosonic form involving 
phonon-like density fluctuation operators $\hat{b}^+_{q \sigma}$ and $\hat{b}_{q \sigma}$. 
The index $\sigma = \pm 1$
accounts for two spin directions or two different components in the Fermi gas.

\item[ii)]
The pair interactions between fermions must be describable in terms of bilinear products of 
density fluctuation operators.

\item[iii)]
In the spinful case, an additional transformation to charge and spin degrees of freedom
(or mass and composition degrees in the neutral case), denoted by $\nu = \pm 1$,  is needed 
on the way to a diagonal Hamiltonian:

\begin{eqnarray}\label{1}
\hat{d}_{q \nu} \equiv \frac{1}{\sqrt{2}} \sum _ \sigma \sigma ^{\frac{1-\nu}{2}} 
\,\hat{b}_{q \sigma},\quad \nu=\pm 1.
\end{eqnarray}

\item[iv)]
In the absence of backscattering between different components, the Hamiltonian then
separates  into decoupled spin and charge (or mass and composition)  parts which can 
be diagonalized by a 
Bogoliubov transformation. The final Hamiltonian $\tilde{H}$ for the excitations of
the one-dimensional Fermi sea typically has the structure
$\tilde{H} = \sum _{q \nu} v_{q \nu}\,|q|\, \hat{f}^+ _{q \nu} \hat{f}_{q \nu}$
in terms of free phonon operators $\hat{f}^+$ and $\hat{f}$ and renormalized velocities
$v_{q \nu}$. Zero modes \cite{Hal81} must be added to make the correspondence with the
original fermionic Hamiltonian complete.

\item[v)]
If the dependence of the velocities on $q$, i.e., that of the underlying coupling functions
is neglected, hermitian phase fields $\hat{\Phi}_\nu(z)$ and dual fields $\hat{\Theta}_\nu(z)$
can be defined 
in terms of the $\hat{d}$-operators (or else in terms of the $\hat{f}$-operators) which contain 
zero modes and bring the Hamiltonian into the canonical form

\begin{eqnarray}\label{3}
 \hat{H} = \sum _\nu \frac{v_\nu}{2} \,\int dz \left [ \pi K_\nu \,
\left(\partial _z \hat{\Theta}_\nu (z)\right)^2 + \frac{1}{\pi K_\nu} \,
\left(\partial _z \hat{\Phi}_\nu(z)\right)^2 \right],
\end{eqnarray}
where $K_\nu$ are the central coupling constants. Backscattering between the two 
components  $\hat{V}_\perp \propto 
\int dz\,\cos\left[\sqrt{8}\,\hat{\Phi}_{-1}(z)\right]$
destroys the simple quadratic form of the spin part \cite{LE74,HSSU75}.

\item[vi)]
An important step in any phase theory is the representation of fermion field operators
in terms of the phase operators in order to calculate correlation functions easily, a procedure 
usually called "bosonization". The
method goes back to \cite{LP74} and \cite{M74} and is now well understood 
\cite{HSU80,Hal81}. For open boundaries, it is described in the works \cite{FG95} 
and \cite{MEJ97}.
\end{itemize}

Here, we consider interacting one-dimensional fermions in a harmonic trap. 
Then the single particle wave functions are not simple combinations of plane waves. We, 
therefore, use a general bosonization
prescription described in \cite{Schoen} which involves auxiliary Fermi fields
$\hat{\psi}_{a \sigma}$ related to the original
operators $\hat{c}_{l \sigma}$ (note that for the harmonic oscillator $q$ is replaced by 
a discrete index $l=...,-2,-1,0,1,2,...$) by

\begin{equation}\label{5}
 \hat{\psi} _{a \sigma} (u) \equiv \sum^\infty _{l=-\infty} e^{i l u}\, \hat{c}_{l \sigma}
= \hat{\psi}_{a \sigma} (u + 2 \pi). 
\end{equation}
These operators are defined on the unit circle as will be most other operators below.
Bosonization is provided by \cite{Schoen,MS00} 

\begin{eqnarray}\label{6}
\hat{\psi} _{a \sigma}(u)=e^{i u \hat{N}_\sigma}\,
e^{i \hat{\phi}^+_\sigma (u)}\,e^{i \hat{\phi}_\sigma (u)}\,\hat{U}_\sigma\
= e^{i u \hat{N}_\sigma}\,
\frac{1}{\sqrt{\eta}}\,e^{i (\hat{\phi}^+_\sigma (u)+ \hat{\phi}_\sigma (u))}
\,\hat{U}_\sigma
\end{eqnarray}
in terms of the bosonization phases

\begin{eqnarray}\label{7}
\hat{\phi} _\sigma (u) = - i \sum _{m=1} \frac{1}{\sqrt{m}}\, e^{ im (u+i \eta/2)}\,
\hat{b}_{m \sigma}.
\end{eqnarray}
The quantity $\eta$ in the non-normal ordered expression is a positive infinitesimal.

There is a nontrivial relation between the bosonization phase $\hat{\phi}^+_\sigma (u)
+\hat{\phi}_\sigma (u)$
and the physical phase operators $\hat{\Phi}_\nu(u)$ and $\hat{\Theta}_\nu(u)$ given 
below. The Klein operator $\hat{U}_\sigma$ reduces the fermion number by one. An explicit
construction of $\hat{k}_\sigma= - i \ln \hat{U}_\sigma$ is given in \cite{Schoen}.
The zero modes in the physical phases $\hat{\Phi}_\nu$ and $\hat{\Theta}_\nu$ are related
to $u \hat{N}_\sigma$ and $\hat{k}_\sigma$ by the transformation equation (\ref{1}).

In the present case of the harmonic trap, the relation between the auxiliary variable $u$
in the correlation functions and the spatial position $z$ inside the trap turns out to be
$u \rightarrow u_0(z) =\arcsin(z/L_F)-\pi/2$,
where $2L_F$ is the quasi-classical extension of the Fermi sea.

The paper is organized as follows: Section 2 develops the phase theory for a one-component 
Fermi gas according to the above prescription. Section 3 extends the theory to two components. 
In Section 4, we calculate the
scaling exponents contained in the one-particle correlation functions. The Appendix gives all
details of their evaluation using the WKB representation developed in \cite{AGW04}.

\section{Phase theory for one component}

We start with the simpler case of one fermionic component, e.g., spin polarized fermionic
atoms in the case of ultracold atoms. 
The one-dimensional equivalent of s-wave scattering, i.e., a contact interaction is 
forbidden by the exclusion principle and the interaction in such a Fermi gas is usually
weak. The model may, nevertheless, have potential application to ultracold Fermi gases 
in a harmonic trap provided Feshbach resonances enhance the interaction as recently found
in the three-dimensional case \cite{R03}. 

Quasi one-dimensionality can be realized
when the particle number $N$ is less than the ratio of transverse to longitudinal
trap frequency $\omega _t/\omega _\ell$ and the three-dimensional interaction energy 
per fermion is less than the transverse excitation energy
in the highly elongated cylindrical trap \cite{GS03}. 

However, it is not enough to reach degeneracy,
$k_B T \ll \epsilon _F$: In order to observe zero temperature behavior, the much stronger
condition $k_B T \ll \hbar \omega _\ell$ is also required. 

The model considered in \cite{WW} is described by the
bosonic excitation Hamiltonian

\begin{eqnarray}\label{8}
\fl \tilde{H}  =  \frac{1}{2} \,\hbar \omega _\ell \, \sum _{m>0} m \left(
 \hat{d}_m \hat{d}_m ^+ +  \hat{d}_m ^+ \hat{d}_m \right)
 + \frac{1}{2} \sum _{m>0} [V_c(m)-V_b(m)]\, \sqrt{2m} \left [ \hat{d}
_{2m} + \hat{d}^+ _{2m} \right ] \\ \nonumber
&&
\\[4mm]\nonumber
+ \frac{1}{2} \sum _{m>0} V_a(m)\, m \left(
 \hat{d}_m \hat{d}_m ^+ +  \hat{d}_m ^+ \hat{d}_m \right)
-\frac{1}{2} \sum _{m>0} [V_c(m)-V_b(m)]\, m
 \left ( \hat{d}_m^2+ \hat{d}^{+2}_m \right).
 \end{eqnarray}
In the case of one component, one does not have to discriminate between $\hat{d}$ and
$\hat{b}$ operators. We treat the one-particle operators from $V_b$ and $V_c$ on the 
same footing to avoid problems with spurious self-energies when $V_b$ is retained.

The interaction terms in equation (\ref{8}) follow uniquely by retaining those parts in the
fermionic pair interaction operator

\begin{eqnarray}\label{9}
\hat{V} =\frac{1}{2} \sum _{mnpq}
V(m,p;q,n)\,(\hat{c}^+_m \hat{c}_q ) ( \hat{c}^+_p \hat{c}_n ),
\end{eqnarray}
which are expressible in terms of density fluctuation operators $\hat{\rho}(m)=\sum _p
\hat{c}^+_{p+m}\hat{c}_p$ (in the basis of harmonic oscillator states $\psi _n(z)$),
 i.e.,

\begin{eqnarray}\label{10}
\fl V (m,p;q,n)  \rightarrow  V _a(|q-m|) \, \delta _{m-q, n-p}+V
_b(|q-m|) \, \delta _{q-m, n-p}
%\\[4mm]\nonumber
+ V _c(|q-p|) \, \delta _{m+q, n+p}.
\end{eqnarray}
In \cite{GW02}, it is shown that the retained matrix elements are dominant in the 
one-dimensional many-fermion system in the harmonic trap. This is related to 
approximate momentum conservation during collisions in the trap.

The interaction matrix elements $V_a(m)$, $V_b(m)$, and $V_c(m)$ correspond to the
Luttinger model coupling functions $g_4(p)$, $g_2(p)$, and $g_1(p)$, respectively. $V_a$
and $V_b$ describe forward scattering and $V_c$ describes $2 k_F$ (backward) scattering.
The notion of backscattering in the one-branch-one-component system is to be interpreted
as follows: The study of the matrix elements showed that the right (left) running part of a
one-particle wave function interferes with the left (right) running part of the wave function 
of another fermion exchanging a momentum of about $2 k_F$.

The drastic reduction of interaction matrix elements necessary to obtain equation 
(\ref{8}) from the full Hamiltonian for a given one-dimensional (effective) pair interaction
between fermions may affect physical details. However, the model will still provide us with 
the correct critical exponents in terms of the central coupling constant $K$. This has been 
explicitly shown in \cite{MS00} for OBC. It must be pointed out that the model 
accounts for the basic interaction mechanisms in one dimension, namely forward and backward 
scattering. It may be more practical to consider $K$ as a parameter taken from experiment, 
as is usually done in applications of the Luttinger model.

\subsection{Phase operators}

The second contribution on the r.h.s. of equation (\ref{8}) represents a one-particle operator
$\hat{V}_1$.
This operator is neglected in the model with open boundary conditions as pointed out 
in \cite{MS00}. The one-particle operator
does not alter the critical behavior, but has quantitative effects on other properties.
$\hat{V}_1$ is exactly taken into account in the present model. For more details, we refer
to \cite{WW}.
 
The Bogoliubov transformation to diagonalize the Hamiltonian in terms of operators $\hat{f}$ 
and $\hat{f}^+$ is given 
by $\hat{d} _m = \hat{f} _m \cosh \zeta _m - \hat{f}^+ _m \sinh \zeta _m+\eta _m
\, \exp(- \zeta _m)$, where the inhomogeneous part
$\eta _{2m} = - [V_c(m)-V_b(m)] \exp(- \zeta _{2m})/(2 \epsilon _{2m}\sqrt{2m})$
originates from $\hat{V}_1$ and differs from zero only for even indices.

The central dimensionless coupling constants $K_{m}$, transformation parameters $\xi _m$,
and  the renormalized level spacings $\epsilon _m$
for the model, equation (\ref{8}), are given by \cite{WW}

\begin{eqnarray}\label{14}
K_m &=& e^{-2 \zeta _m} =\sqrt{\frac{\hbar \omega _\ell +
V_a(m) - (V_b(m)-V_c(m))}{\hbar \omega _ \ell + V_a(m) + (V_b(m)-
V_c(m))}}\equiv \frac{K'_m}{2 \pi}, \quad
\\ [4mm] \nonumber
\epsilon _m &=& \sqrt{(\hbar \omega _\ell+V_a(m) )^2 -
(V_b(m)-V_c(m))^2},
\end{eqnarray}
respectively.

In the phase formulation, the dependence of $K$ and $\epsilon$ on $m$ must be suppressed.
Note that $V_a(m) \rightarrow V_a$ implies that there are no interaction effects due to
this matrix element in the one component theory and $V_a$ should be omitted, i.e.,
we use $\epsilon = \sqrt{(\hbar \omega _\ell )^2 - (V_b-V_c)^2}$ and

\begin{eqnarray}\label{15b}
K'=2 \pi K = 2 \pi\,e^{-2 \zeta} = 2 \pi\,\sqrt{\frac{\hbar \omega _\ell     
- (V_b-V_c)}{\hbar \omega _ \ell  + (V_b- V_c)}},\quad \epsilon=\hbar \omega _\ell\,
\frac{2 K}{K^2+1},
\end{eqnarray}
where $V_b$ is usually small in comparison to $V_c$, even in the marginally long-ranged case 
of the dipole-dipole interaction \cite{GW02}.

In some physical quantities, e.g., the one-particle correlation function, neglect of the 
dependence on $m$ in $K$ (which for large $m$
must approach unity \cite{Hal81}) leads to divergences. We take care of this by writing

\begin{eqnarray}\label{15c}
K_m=1+(K-1)\,\exp(-mr),
\end{eqnarray}
with a small positive number $r \ll 1$, estimated as $r \approx R/L_F \propto 1/ \sqrt{N}$ 
where $R$
is the range of the interaction. This small number should be discriminated from the positive
infinitesimal $\eta$ appearing in equation (\ref{6}). For $m < 1/r \gg 1$, there is 
practically no dependence on $m$. 

We begin with rewriting the one-particle operator $\hat{V}_1$. This already leads to the
structure of the main phase operator.
Assuming the same exponential decay as in equation (\ref{15c}) also for interaction matrix 
elements, i.e., $V_c(m)-V_b(m)= (V_c-V_b) \,\exp(-m r)$, $\hat{V}_1$ can be expressed as

\begin{eqnarray}\label{16}
\hat{V}_1=\frac{1}{4 \pi}\,(V_ c-V_b) \,\int _ {-\pi}^\pi du\,\left[\frac{e^{-r +2 i u}}
{1-e^{-r+2 i u}}
+\frac{e^{-r -2iu}}{1-e^{-r -2 i u}}\right]\,\partial _u \hat{\phi}_{odd}(u),
\end{eqnarray}
where

\begin{eqnarray}\label{16a}
\fl \hat{\phi}_{odd}(u) \equiv \frac{1}{2}\,(\hat{\phi}(u)+\hat{\phi}^+(u)-\hat{\phi}(-u)-
\hat{\phi}^+(-u))
= \sum _{n=1}^\infty \sqrt{\frac{e^{-n \eta}}{n}}\, 
\,\sin(n u)\,\left ( \hat{d}_n +\hat{d}^+ _n \right).
\end{eqnarray}
The same phase operator appears in the particle density operator when it is decomposed
into a slowly varying part and a part describing Friedel oscillations \cite{AGW04}.

At the level of phase operators, the presence of $\hat{V}_1$ causes a c-number shift
 
\begin{eqnarray}\label{17}
b(u)=i \,\frac{K (V_c-V_b)}{4 \epsilon }\,\ln\left(\frac{1-e^{-r+2 i u}}
{1-e^{-r -2 i u}} \right)\equiv i \kappa _0 \,\ln\left(\frac{1-e^{-r+2 i u}}
{1-e^{-r -2 i u}} \right),
\end{eqnarray}
with $\kappa _0=(K^2-1)/8$. The main phase operator $\hat{\Phi}(u)$ corresponds to that in 
\cite{MEJ97}:

\begin{eqnarray}\label{18} 
\fl \hat{\Phi}(u) \equiv \hat{N}u+\hat{\phi}_{odd}(u)+b(u)= \hat{N}u+\sum _{m=1}^\infty 
\sqrt{\frac{K_m\,e^{-m \eta}}{m}}
\,\sin (m u)\, (\hat{f}_{m}+\hat{f}^+_{m}).
\end{eqnarray}
Note, that in \cite{MEJ97} rescaled phases are used and the function $b(u)$ is absent due to
the neglect of one-particle operators. The dual phase operator is

\begin{eqnarray}\label{19} 
\fl \hat{\Theta}(u) \equiv \frac{i}{2 \pi}\, \sum _{n=1}^\infty \sqrt{\frac{\,e^{-n \eta}}{n}} 
\,\cos(n u)\,\left (\hat{d}_{n} -\hat{d}^+ _{n} \right)
=\frac{i}{2 \pi}\, \sum _{n=1}^\infty \sqrt{\frac{e^{-n \eta}}{n K_n}}\, 
\cos(n u)\,\left (\hat{f}_{n} -\hat{f}^+ _{n} \right).
\end{eqnarray}

We prefer not to combine the zero mode operator $\hat{k}$ with the dual phase field 
$\hat{\Theta}$. Instead, 
we retain the unitary operator $\hat{U}=\exp(i \hat{k})$ in order to avoid the mathematical 
problems pointed out in \cite{Schoe01}. The relevant commutator is 
$[ \exp(i \hat{k}),\hat{N}]=\exp(i \hat{k})$. 
The phase operator corresponding to the momentum density is

\begin{eqnarray}\label{21} 
\hat{\Pi}(u) &\equiv& \partial _u \hat{\Theta}(u)= - \frac{i}{2 \pi}\sum _{n=1}^\infty 
\sqrt{n\,e^{-n \eta}}
\,\sin(n u)\,(\hat{d}_n - \hat{d}^+_n)
\\[4mm]\nonumber
&=&- \frac{i}{2 \pi}\sum _{n=1}^\infty 
\sqrt{\frac{n\,e^{-n \eta}}{K_n}}
\,\sin(n u)\,(\hat{f}_n - \hat{f}^+_n).
\end{eqnarray}
Using 
$\sum _{m=1}^\infty \left(e^{im(u+i \eta/2)}+e^{-im(u-i \eta/2)}\right)\rightarrow 2\pi\,
\delta _{2 \pi}(u)-1$ leads to the commutator

\begin{eqnarray}\label{23}  
 \left [ \hat{\Phi}(u), \hat{\Pi}(v)  \right]
 = \frac{i}{2} \left(\delta _{2 \pi}(u-v) - \delta _{2 \pi} (u+v) \right), 
 \end{eqnarray}
which is also implicit in the work \cite{MEJ97}.
This is not a canonical commutator. Canonical commutation relations cannot hold for 
operators with fixed parity. 

\subsection{Phase Hamiltonian}

Inserting equation (\ref{18}) and  equation (\ref{21}) into

\begin{eqnarray}\label{24} 
\hat{H}= \frac{\epsilon}{2}\, \int ^\pi _{-\pi} \, du \left [ K'\,
\hat{\Pi}^2(u) + \frac{1}{K'} \left (\partial _u \hat{\Phi}(u)\right)^2  \right]
=\int _{- \pi}^\pi du \, \hat{{\cal{H}}},
\end{eqnarray}
and neglecting an irrelevant contribution from $b^2(u)$, 
reproduces  equation (\ref{8}) in addition to the zero mode contribution $E_0(\hat{N})$

\begin{eqnarray}\label{25} 
E_0(\hat{N})=\frac{\epsilon}{2 K}\,\hat{N}^2=\frac{\hbar \omega _\ell}
{K^2+1}\,\hat{N}^2.
\end{eqnarray}
In the latter relation, corrections $O(\hat{N})$ are neglected.
Remarkably, canonical equations still hold in spite of  equation (\ref{23}): Consider the 
Heisenberg phase field $\hat{\Phi} (u,t)$ (we set $\hbar=1$)

\begin{eqnarray}\label{26}
 i \partial _t \hat{\Phi} (u,t) = \left [ \hat{\Phi}(u,t), \hat{H}(t)  \right]
=i \epsilon \,K'\, \hat{\Pi}(u,t) \equiv i \frac{\partial \hat{{\cal{H}}}}{\partial \hat{\Pi}
(u,t)}.
 \end{eqnarray}
Thus $\partial _t \hat{\Phi}(u,t) = \epsilon \,K'\,\hat{\Pi}(u,t)$ remains correct.

\subsection{Bosonization}

A simple expression of the individual operators $\hat{\phi}$ and $\hat{\phi}^+$ in  
equation (\ref{6})
in terms of the new phase operators $\hat{\Phi}$ and $\hat{\Theta}$ is not available. 
However, their sum which appears in the non-normal ordered form of  equation (\ref{6}),
is represented by

\begin{eqnarray}\label{31}
\hat{\phi}(u)+\hat{\phi}^+(u)=\hat{\phi}_{odd}(u)-2 \pi\,\hat{\Theta}(u) \equiv
\hat{\Phi}(u)-b(u)-\hat{N}u -2 \pi\,\hat{\Theta}(u).
\end{eqnarray}
This can be utilized to calculate the auxiliary correlation function

\begin{eqnarray}\label{32}
 \langle \hat{\psi}^+ _a (u,t)\psi _a (v)\rangle  = e^{-i(N-1)(u-v)+ i \mu _N t} \,
 \frac{1}{\eta}\langle e^{-i \hat{\phi}^+ (u,t)-i \hat{\phi}(u,t)}
e^{i \hat{\phi}^+ (v)+i \hat{\phi}(v)} \rangle,
\end{eqnarray}
with the chemical potential

\begin{eqnarray}\label{33}
\mu _N \equiv E_0(N)-E_0(N-1)=\frac{2 \hbar \omega _\ell}{K^2+1}\,\left(N-\frac{1}{2}\right).
\end{eqnarray}
Analogously

\begin{eqnarray}\label{32a}
 \langle \hat{\psi} _a (u,t)\hat{\psi}^+ _a (v)\rangle  = e^{iN(u-v)- i \mu _{N+1} t} \,
 \frac{1}{\eta}\langle e^{i \hat{\phi}^+ (u,t)+i \hat{\phi}(u,t)}
e^{-i \hat{\phi}^+ (v)-i \hat{\phi}(v)} \rangle.
\end{eqnarray}
Derivation and evaluation of these expressions is described in the Appendix.

\section{Phase Theory for two components}

In terms of mass and composition operators, i.e., after the transformation  equation (\ref{1}),
the excitation Hamiltonian for forward scattering is given by 

\begin{eqnarray}\label{34}
\tilde{H}_{for} &\equiv& \frac{1}{2} \sum _{m \nu} m [ \hbar \omega _\ell + V _{a \parallel}
(m) + \nu V _{a \perp} (m) ] \left \{ \hat{d} ^+ _{m \nu} \hat{d}_{m \nu} + \hat{d}_{m \nu}
\hat{d}^+ _{m \nu} \right \} 
\\ \nonumber
&&+ \frac{1}{2} \sum _{m, \nu} m [ V_ {b \parallel} (m) + \nu V_{b \perp} (m)]
\left \{ \hat{d}^{+2}_{m \nu} + \hat{d}^2_{m \nu} \right \}. 
\end{eqnarray}
It must be supplemented by backward scattering and one-particle contributions. Noting that the
one-particle operator
$\frac{1}{2} \sum _{m >0,\sigma} \sqrt{2m}\, (V_{c \parallel}(m)-V_{b \parallel}(m))\,
[ \hat{b} _{2m \sigma} + \hat{b}^+ _{2m \sigma}]$,
which originates from rearranging the two-particle backscattering operator and from a spurious 
self-energy in $\hat{V}_b$, is transformed by means of $\sum _\sigma \sigma ^{ \frac{1-\nu}{2}} 
\equiv 1 + \nu= 2\delta _{\nu,1}$ and

 \begin{eqnarray}\label{37}
 \fl \sum _\sigma \,[ \hat{b}_{m \sigma} + \hat{b}^+ _{m \sigma}] =
 \frac{1}{\sqrt{2}} \sum _\nu \left ( \sum _ \sigma \sigma^{\frac{1-\nu}{2}} \right)\,
 \left ( \hat{d}_{m \nu} + \hat{d}^+ _{m \nu} \right)
 = \sqrt{2}\left( \hat{d}_{m 1} + \hat{d}^+ _{m 1} \right),
 \end{eqnarray}
 the final form of the additional operator is
 
\begin{eqnarray}\label{38}
\fl\hat{V}_{add}=  
-\frac{1}{2}\sum _{m>0, \nu} m  V _{c \parallel} (m) \left \{
\hat{d}^{+2}_{m \nu} + \hat{d}^2_{m \nu} \right \}
+ \sum _{m>0} \sqrt{m}[V_{c \parallel}(m)-V_{b \parallel}(m)]
\left (\hat{d}^+_{2 m 1} + \hat{d}_{2 m 1} \right).
\end{eqnarray}
 
So far, we did not mention $ V_{c \perp}$, i.e., backscattering between the two components, 
but we note that $ V_{c \perp}$ does not generate a one-particle potential. 

In omitting the dependence on m, $V _{a \parallel}$ drops out again. Then the
renormalized level spacings and central coupling constants are given by
 
 \begin{eqnarray}\label{40}
 \epsilon _{\nu} = \sqrt{(\hbar \omega _\ell + \nu V_{a \perp})^2 -
[V_{b \parallel} + \nu V_{b \perp}- V_{c \parallel}]^2},
\end{eqnarray}
and

\begin{eqnarray}\label{41}
K _{\nu} = \sqrt{\frac{(\hbar\omega _\ell + \nu V_{a \perp})    
-[ V_{b \parallel}+ \nu V_{b \perp}-V_{c \parallel}]}{(\hbar \omega _ \ell  
+ \nu V_{a \perp})+ [V_{b \parallel}+\nu V_{b \perp}- 
V_{c \parallel}]}} \equiv \frac{K' _\nu}{2 \pi},
\end{eqnarray}
respectively. The total one-particle operator from $V_b$ and $V_c$ can be rewritten as

\begin{eqnarray}\label{42}
\fl \hat{V}_1=\frac{1}{4 \pi \sqrt{2}}\,(V_ {c \parallel}-V_{b \parallel})\, \sum _{\nu}(1+\nu)
\,\int _ {-\pi}^\pi du\,\left[\frac{e^{-r +2 i u}} {1-e^{-r+2
i u}} +\frac{e^{-r -2iu}}{1-e^{-r -2 i u}}\right]\,\partial _u
\hat{\phi}_{\nu,odd} (u), 
\end{eqnarray}
with

\begin{eqnarray}\label{43}
\hat{\phi}_ {\nu,odd} (u) \equiv  \sum _{n=1}^\infty \frac{1}{\sqrt{n}}\,e^{-n \eta/2}\, 
\sin (nu)\left  ( \hat{d}_{n \nu}+ \hat{d}^+ _{n \nu} \right).
\end{eqnarray}
The corresponding phase shifts are the c-number functions (real and odd) 

\begin{eqnarray}\label{44}
\fl b_{\nu}(u)=i \,\frac{K'_{\nu} (V_{c \parallel}-V_{b \parallel})}{8 \pi \epsilon _{\nu} 
\sqrt{2}}(1+\nu)\,
\ln\left(\frac{1-e^{-r+2 i u}}{1-e^{-r -2 i u}} \right)\equiv i \kappa _1\,\delta _{\nu,1}
\ln\left(\frac{1-e^{-r+2 i u}}{1-e^{-r -2 i u}} \right),
\end{eqnarray}
with $\kappa _1 \equiv (K^2_1-1)/8$. Because of $b_{-1}=0$ only the mass part
is affected by the one-particle operator. The phase operators are

\begin{eqnarray}\label{45}
\fl\hat{\Phi}_\nu (u) \equiv \hat{\phi}_{\nu,odd}(u)+b_\nu(u)+u \hat{N}_\nu
=\sum _{n=1}^\infty \sqrt{\frac{K_{n \nu}}{n}}\,\,e^{-n \eta/2}\,\sin (nu)\,(\hat{f}_{n \nu}
+\hat{f}^+_{n \nu}) +u \hat{N}_\nu,
\end{eqnarray}
with $\hat{N}_\nu=\sum _ \sigma \sigma^{\frac{1-\nu}{2}}\,\hat{N}_\sigma/\sqrt{2}$
and the dual phase operator

\begin{eqnarray}\label{46}
\fl \hat{\Theta}_{\nu} (u) \equiv \frac{i}{2 \pi}\, \sum _{n=1}^\infty \sqrt{\frac{e^{-n \eta}}
{n}}\,
\cos(n u)\,\left(\hat{d}_{n \nu} -\hat{d}^+ _{n \nu} \right)
=\frac{i}{2 \pi}\,\sum _{n=1}^\infty \sqrt{\frac{e^{-n \eta}}{nK_{n \nu}}}\,\cos (nu)\,
(\hat{f}_{n \nu} -\hat{f}^+_{n \nu}). 
 \end{eqnarray}
The dual phase fields $\hat{\Theta}_\nu$ generate associated momentum densities via

\begin{eqnarray}\label{47}
 \hat{\Pi}_{\nu} (u) \equiv \frac{\partial}{\partial
u}\,\hat{\Theta}_{\nu}(u) = - \frac{i}{2 \pi}\, \sum _{n=1}^\infty \sqrt{ne^{-n \eta}}
\,\sin (n u) \left(\hat{d}_{n \nu}- \hat{d} _{n \nu}^+\right) .
\end{eqnarray}
Finally, the phase Hamiltonian is

\begin{eqnarray}\label{49}
 \hat{H} = \sum _ \nu \frac{\epsilon _\nu}{2} \int ^\pi
_{- \pi} du \left \{K'_\nu \hat{\Pi}_{\nu}^2 (u)+\frac{1}{K'_{\nu}}\left (
\partial _u \hat{\Phi}_{\nu} (u)  \right)^2\right\}.
\end{eqnarray}
The zero mode in $\hat{\Phi}_\nu$ gives the ground state energy operator

\begin{eqnarray}\label{50}
\hat{E}_{0 \nu}=\frac{\epsilon _\nu}{4 K_\nu}\left(\delta _{\nu, 1}\sum
_{\sigma \sigma'} \hat{N}_\sigma\,\hat{N}_{\sigma'}
+\delta _{\nu, -1}\left(\sum _\sigma \sigma \hat{N}_\sigma \right)^2 \right).
\end{eqnarray}
Again, the equations of motion for the Heisenberg fields $\hat{\Phi}_\nu(u,t)$ are
$\partial _t \hat{\Phi}_\nu(u,t) = \epsilon _\nu\,K'_\nu\, \hat{\Pi}_\nu(u,t)$.

Bosonization according to equation (\ref{6}) is provided by 

\begin{eqnarray}\label{53}
 \hat{\phi}_\sigma(u)+\hat{\phi}^+_\sigma(u)=\sum _\nu \frac{\sigma^{\frac{1-\nu}{2}}}
{\sqrt{2}}\,(\hat{\Phi}_\nu(u)-b_\nu(u)-u \hat{N}_\nu -2 \pi\,\hat{\Theta} _\nu(u)). 
\end{eqnarray}

Backscattering between different components is described by the operator 
$\hat{V}_{c \perp}$. The composition 
part $\hat{H}_{\nu=-1}$ acquires the form of a sine-Gordon Hamiltonian on the unit circle.
Considering the results for the usual sine-Gordon system, one can speculate that a
sufficiently repulsive interaction renders $\hat{V}_{c \perp}$ irrelevant at low energies.
Restricting the treatment to long range interactions as discussed in \cite{MS00} avoids this 
problem.

\section{Critical exponents}

This section compiles the results for critical exponents extracted from the one-particle correlation 
functions which are calculated in the Appendix. Subsections are devoted to the static 
two-point correlation function, the local dynamic correlation function, and to the local 
spectral density of states, respectively.

\subsection{Static two-point correlation function}

We consider the zero temperature static correlation function

\begin{eqnarray}\label{54}
C(z_1,z_2)\equiv C(z_1,t=0;z_2)\equiv\langle \hat{\psi}^+(z_1)\,\hat{\psi}(z_2) \rangle _0.
 \end{eqnarray}
 We have to evaluate  equation (\ref{A19}) using the explicit result (\ref{A28}) 
for $t=0$. To this order, we introduce the abbreviations 
$Z _\nu \equiv (1-z^2_\nu/L_F^2)^{1/2};\quad \nu = 1,2$, $L_F=\ell \sqrt{2N-1}$ 
($\ell$ is the oscillator length)
to find for $L_F-|z_\nu| > L_F r^2$ and $z_1 \neq z_2$ (the case $z_1=z_2$ was 
considered in \cite{AGW04})
  
\begin{eqnarray}\label{56}
\fl |{\cal{D}} (\pm 2 u_0(z_\nu)+ir) |^2 \rightarrow \frac{1}{4 Z _\nu ^2},\,
    |{\cal{D}}(u_0 (z_1)\pm u_0(z_2)+ i r)|^2 \rightarrow \frac{1}{2 \left(1 \pm Z_1
 Z_2 - z_1 z_2/L_F^2\right )}.
 \end{eqnarray}
 This gives
 
 \begin{eqnarray}\label{58}
\fl C(z_1, z_2)= \frac{r^{2 \gamma _0}2 ^{\alpha _0 + \frac{1}{2}- \gamma _0}}
 {2 \pi L_F} 
\Bigg [  \frac{ \sin (k_F(\tilde{z}_1- \tilde{z}_2)+ A_-)}
 {(1-Z_1 Z_2- z_1 z_2/L^2_F)^{\gamma _0 + \frac{1}{2}}(1+Z_1 Z_2-
 z_1 z_2/L^2_F)^{\alpha _0}}
 \\[4mm]\nonumber
 \fl + (-1) ^{N-1} \frac{\cos (k_F (\tilde{z}_1+\tilde{z}_2)+A_+)}
 {(1+Z_1 Z_2 - z_1z_2/L^2_F )^{\gamma _0 + \frac{1}{2}}
 ( 1 - Z_1 Z_2- z_1z_2/L^2_F) ^{\alpha _0}} \Bigg]\,(Z_1 Z_2)^{\alpha _0 
 - \frac{1}{2}}, 
 \end{eqnarray}
 with the (irrelevant) phase shift
$ A_ \pm \equiv \frac{1}{2}\left ( \arcsin (z_1/L_F) 
 \pm \arcsin (z_2/L_F) \right) -b(u_0(z_1)) \mp b(u_0(z_2))$ 
 and notations from the Appendix, e.g.,
$ \tilde{z}_\nu  = \tilde{z}(z_\nu)= z_\nu Z(z_\nu)/2 
 + (L_F/2)\,  \arcsin (z_\nu/L_F)$.
 For $z_1 \equiv z >\ell$ and $z_2=0$, and well inside the trap ($|z| \ll L_F$), we find
 
 \begin{eqnarray}\label{61}
C_{\mbox{\small center}} (z) = \frac{r^{2 \gamma _0}}{\pi L_F} \, \frac{\sin k_F z}
 {(z/L_F)^{2 \gamma _0+1}}+ (-1)^{N-1} \frac{r^{2 \gamma _0}}{2 \pi L_F}
 2^{2 \alpha _0 - 2 \gamma _0} \frac{\cos k _F z}{(z/L_F)^{2 \alpha _0}}. 
 \end{eqnarray}
 Because of $ |z|/L_F \ll 1 $ and $ 2 \gamma _0 + 1 = 2 \alpha _0 + K$,
 the second term is not important. Thus
 
 \begin{eqnarray}\label{62}
 C_{\mbox{\small center}} (z) = \frac{r^{2 \gamma _0}}{\pi L_F}\, \frac{\sin k_F z}
 {(z/L_F)^{2 \gamma _0+1}},
 \end{eqnarray}
 gives the decay exponent (twice the bulk scaling dimension of $\hat{\psi}$)
 
 \begin{eqnarray}\label{63}
\alpha _C = 2 \gamma _0 + 1= \frac{1}{2}\,\left[K+\frac{1}{K}\right]\equiv 2 \Delta.
 \end{eqnarray}
 This result was obtained numerically for the present model in the second paper of \cite{WW}. 
 It coincides 
 not unexpectedly with the OBC result \cite{FG95,MEJ97,MS00} in the bulk limit and also 
 with that for the homogeneous Luttinger model. 
 
 We can extract from  equation (\ref{58}) the boundary exponent $\Delta _\perp$. To this 
 order, the thermodynamic limit must be performed explicitly: 
 As in \cite{DG99} this is done by making the trap shallower and shallower: $\omega _\ell
 \propto 1/N \rightarrow 0$. Furthermore, the coupling constant $K$ and the Fermi wave 
 number $k_F$ must be held fixed. For a specific interaction model, this limit was 
 already performed in \cite{AGW04}. Furthermore, besides $\eta$ also $r \propto
 1/\sqrt{N}$ vanishes and one should consider renormalized fields as described in 
 \cite{Sh95}.
 
 Consider then the boundary situation $z_2 =L_F(1-\eta)$ while $0 < z_1 \ll L_F$,
 such that $z_2 \rightarrow L_F$ that is $Z_2 \rightarrow 0$ in the denominator of 
 $C(z_1,z_2)$. The distance $z_2-z_1$ then is large as required for $\Delta _\perp$.
  Equation (\ref{58}) gives
 
 \begin{eqnarray}\label{65} 
\fl C(z_1,L_F(1-\eta)) \propto \frac{Z_1^{\alpha _0-1/2}}{(1- z_1/L_F)^{\alpha _0 + \gamma _0 
 + \frac{1}{2}}}\,(\sqrt{\eta})^{\alpha _0-1/2}\propto \frac{(z_2-z_1)^{(\alpha _0-1/2)/2}}
 {(z_2-z_1)^{\alpha _0 + \gamma _0 + \frac{1}{2}}}. 
 \end{eqnarray}
 Hence
 
 \begin{eqnarray}\label{66}
 2 \Delta _\perp = \frac{1}{2}\,\alpha _0 + \gamma _0 + \frac{3}{4} = \frac{1}{8}\,
 \left [ \frac{3}{K} + K +2 \right].
 \end{eqnarray}
 This boundary exponent is different from the corresponding result for OBC in \cite{FG95,MEJ97}. 

\subsection{Local dynamic correlation function}

Following \cite{MEJ97,MS00}, we consider only the part varying slowly in space, i.e., we
neglect the rapidly oscillating Friedel part given by the last two terms in  equation (\ref{A19}). 
We then obtain

\begin{eqnarray}\label{67}
C^{NF}(z,t;z) = \frac{1}{\pi L_F Z(z)} \langle \hat{\psi}_a^+ (u_0(z),t) 
\hat{\psi}_a (u_0(z)) \rangle _0.
\end{eqnarray}
According to  equation (\ref{A28}), 

\begin{eqnarray}\label{68}
\langle \hat{\psi}^+ _a (u_0,t)\hat{\psi}_a (u_0)\rangle _0 = e^{i \mu _N t} \,r ^{2 \gamma _0} 
 \, \left[{\cal{D}}(-\epsilon t + ir)^{2 \gamma _0}\,{\cal{D}}(-\epsilon t + i\eta) \right]
 \\[4mm]\nonumber
\times   \left\{\frac{ {\cal{D}}(2u _0-\epsilon t + i r)\,{\cal{D}}(-2 u_0-\epsilon t+i r)}
   {|{\cal{D}}(2u_0+ir)|^2}\right\}^{\alpha _0}  
\end{eqnarray}
holds with $u_0=\arcsin(z/L_F)-\pi/2$.
The renormalized level spacing $\epsilon $ vanishes in the thermodynamic limit, but the 
chemical potential becomes a constant independent of particle number:
 $\mu _N=\epsilon\,(N-1/2)/K \rightarrow  2 \hbar/(K^2+1)\,(\omega _\ell N) 
=const.$. Vanishing $\eta$ and $r$ imply

\begin{eqnarray}\label{70}
\fl {\cal{D}}(- \epsilon t) = -\frac{i e^{i \epsilon t/2}}{2 \sin
 (\epsilon t/2)}, \quad {\cal{D}}(2u_0-\epsilon t)
 {\cal{D}}(-2u_0-\epsilon t) = \frac{e^{i \epsilon t}}
 {4 \left [ \cos ^2( \epsilon t/2)- z^2/L_F^2  \right]}. 
 \end{eqnarray}
In order to collect all possible powers of $t$, we specify $z$ to be so near the classical 
boundaries that $\cos ^2( \epsilon t/2)- z^2/L_F^2 \propto t^2$ holds. Then we obtain:

\begin{eqnarray}\label{71}
C^{NF} \propto {\cal{D}} (-\epsilon t) ^{2 \gamma _0 + 2 \alpha _0 + 1}\propto
 \frac{1}{t^ {2 \Delta _\parallel}},
\end{eqnarray}
with the boundary exponent

\begin{eqnarray}\label{72} 
2 \Delta _{\parallel} = 2 \gamma _0 + 2 \alpha _0 + 1 = \frac{1}{K}.
\end{eqnarray}
This agrees with the one-component results in \cite{FG95,MEJ97,MS00} for OBC.
 
However, for $z$ away from the boundaries, only ${\cal{D}}(-\epsilon t)^{2 \gamma _0+1}$
contributes and the bulk scaling exponent is reobtained
 
 \begin{eqnarray}\label{73}
2 \gamma _0 + 1 = \frac{1}{2}\,\left[K+\frac{1}{K}
 \right] \equiv 2 \Delta.
\end{eqnarray} 
 
Note that the scaling relation $2 \Delta _\perp = \Delta +\Delta _\parallel$ \cite{LuR75} 
is not fulfilled with respect to the variables $z$ and $t$. However, boundary conformal 
field theory \cite{Car84} applies to the auxiliary model which uses the variables $\epsilon 
\tau$ ($it \equiv \tau$) and $u$: The infinite strip $w \equiv \epsilon \tau
- i u$ with $-\pi \le u \le 0$ can be mapped conformally onto the complete complex plane
and the Euclidean Lagrange density derived from equation (\ref{24}) is locally rotation 
invariant. Consequently, the scaling relation holds irrespective of the boundary condition.
This can be checked by direct calculation: If we calculate 
$\langle \hat{\psi}^+ _a (u \approx -\pi/2 )\hat{\psi}_a (v \approx 0, \pi)\rangle _0 $ 
instead of the physical correlation function $C(z_1 \approx 0,z_2 \approx L_F, -L_F)$ in 
order to get the boundary exponent $\Delta ^{(a)}_\perp$ for the auxiliary fields we find 
 
\begin{eqnarray}\label{73a}   
\langle \hat{\psi}^+ _a (u)\,\hat{\psi}_a (v)\rangle _0 \propto
\frac{1}{\left|\sin(\frac{u-v}{2})\right|^{2 \Delta ^{(a)}_\perp}},
\end{eqnarray}
with the new value 

\begin{eqnarray}\label{73b}
2 \Delta ^{(a)}_\perp = \frac{1}{4}\,\left[K+\frac{3}{K}\right].
\end{eqnarray}
The critical exponents $\Delta$ and $\Delta _\parallel$ remain unchanged so that the 
scaling relation is fulfilled. 

The subsequent transformation to the physical plane $z_c \equiv \bar{z}+ i \epsilon \tau$
($\bar{z}\equiv z/L_F$) which must ensure $\bar{z}=\cos u$ is, however, not conformal.
The breakdown of local conformal invariance
with respect to the scaled coordinates $\bar{z}$ and $\epsilon \tau$ near the boundaries 
becomes obvious by inspection of the corresponding Euclidean Lagrange density 

\begin{eqnarray}\label{73c}
\hat{{\cal{L}}}_E = \frac{\epsilon}{2 K'} \left \{\left(\partial _{\epsilon\tau} \hat{\Phi}
\right)^2+ Z^2 \left(\partial _{\bar{z}} \hat{\Phi}\right)^2 \right\},  
\end{eqnarray}
with $Z^2=1-\bar{z}^2$.

\subsection{Local spectral density}   
   
We must evaluate the anti-commutator omitting the Friedel part

\begin{eqnarray}\label{74}
 A(t,z) \equiv \left\langle \left [ \hat{\psi}^+ (z,0),\hat{\psi}(z,t) \right] _+ 
\right\rangle^{NF}_0. 
\end{eqnarray}
Using  equation (\ref{A28}) and  equation (\ref{A30}) of the Appendix, we find

 \begin{eqnarray}\label{75}
\fl A(t,z)= \Bigg(e^{-i \mu _{N+1}t} 
{\cal{D}}(-\epsilon t + ir)^{2 \gamma _0}{\cal{D}}(- \epsilon t+i\eta) 
\left\{\frac{{\cal{D}}(-2u_0 - \epsilon t + ir){\cal{D}}(2u_0-\epsilon t + ir)}
{|{\cal{D}}(2u_0)|^2}\right\}^{\alpha _0}
 \\[4mm]\nonumber
\fl + e ^{-i \mu _{N}t} \,{\cal{D}}(\epsilon t + ir)^{2 \gamma _0}\,{\cal{D}}(\epsilon t
  +i \eta) \left \{ \frac{{\cal{D}}(-2u_0 + \epsilon t + ir){\cal{D}}(2u_0+\epsilon t + ir)}
 {| {\cal{D}}(2u_0)|^2}  \right\}^{\alpha _0}\Bigg)\,\frac{r^{2 \gamma _0}}{ \pi L_F Z}. 
\end{eqnarray}
 In the thermodynamic limit, $ r \rightarrow 0$, $\mu _N \rightarrow \mu _{N+1}\rightarrow 
 \mu$, and measuring the frequency $\omega$ from the chemical potential, $\omega-\mu \rightarrow
 \omega$, we obtain
 
 \begin{eqnarray}\label{76}
\fl  N(\omega, z)= \frac{1}{2 \pi}\,\int _{-\infty}^\infty dt \,e^{i \omega t}\,A(t,z)
\propto \frac{Z^{2 \alpha _0-1}(z)2^{-2 \gamma _0-1}}{\pi ^2 L_F} 
\\[4mm]\nonumber
\times\int^\infty _0 \, dt \cos (\omega t) \left \{ \frac{e^{-i(\gamma _0 + 1/2)
\pi + i (\gamma _0 + 1/2+ \alpha _0)\epsilon t}}{\sin ^{2 \gamma _0+1}
(t \epsilon/2)\left [ \cos ^2 (t \epsilon/2) - z^2/L_F^2 \right]
^{\alpha _0} } +c.c.\right\}.  
\end{eqnarray}
 We further proceed as in \cite{MEJ97}: The regularization $\cos (\omega t) \rightarrow 
 [\cos(\omega t)-1]$ is adopted and the integral
$ \int ^\infty _0 \, dt [ \cos(t) -1]/t^k= \pi/(2 \Gamma (k)\cos(\pi k/2)),\quad
 1 < k< 3$ is used. Well inside the trap, we find
 
 \begin{eqnarray}\label{78}
 N (\omega, |z| \ll L_F) \propto \frac{1}{\pi L_F }\, \frac{1}{\epsilon \Gamma (2 \gamma _0+1)}  
 \left ( \frac{\omega}{\epsilon}  \right ) ^{2 \gamma _0}.  
\end{eqnarray}
The bulk exponent of the local spectral density thus is
 
\begin{eqnarray}\label{79}
 \alpha _{\mbox{\small bulk}} = 2 \gamma _0=\frac{1}{2}\,\left[K+\frac{1}{K}-2 \right],\quad 0 < 
 \gamma _0 <1,
\end{eqnarray}
and agrees with the corresponding result in \cite{MEJ97,MS00} for OBC. 

The boundary exponent
is obtained as follows: Near a boundary, $|z| \approx L_F$, we obtain for  
$4 Z^2 \ll \epsilon^2 t^2 \ll 1$, when $\left [ \cos ^2( t \epsilon/2)- z^2/L_F^2  
\right] ^{\alpha _0}\rightarrow  e^{i \pi \alpha _0} 2^{-2 \alpha _0} (t \epsilon)^
{2 \alpha _0}$,

 \begin{eqnarray}\label{80}
 N(\omega, |z| \approx L_F) \propto\frac{1}{\pi L_F} \frac{2^{2 \alpha _0}\,
 Z^{2 \alpha _0-1}(z)}{\epsilon \,\Gamma (2 \gamma _0
 +2 \alpha _0+1)} \left ( \frac{\omega}{\epsilon} \right)^{2 \gamma _0 + 2 \alpha _0}.
 \end{eqnarray}
 The boundary critical exponent thus is

\begin{eqnarray}\label{81}
\alpha _{\mbox{\small boundary}} = 2 \gamma _0 + 2 \alpha _0 = \frac{1}{K}-1,\quad 0<\alpha _0
+\gamma _0 <1.
\end{eqnarray}
Again, this value agrees with the one-component version extracted from \cite{FG95,MEJ97,MS00}
for OBC.

\subsection{Two components}

It is straightforward to extend these results to the case of two components: The correlation 
function factorizes in the mass and composition representation. However, the corresponding 
exponents in  equation (\ref{A19a}) have half the previous value due to the transformation
 equation (\ref{1}). This leads to 

\begin{eqnarray}\label{82}
 2 \Delta = \frac{1}{4}\, \left [{K}_1 + K_{-1} +\frac{1}{K}_1+\frac{1}{K}_{-1} \right],\quad
2 \Delta _\parallel= \frac{1}{2}\, \left [\frac{1}{K}_1+\frac{1}{K}_{-1} \right],
\\[4mm]\nonumber
2 \Delta _\perp= \frac{1}{16}\, \left [{K}_1 + K_{-1} +\frac{3}{K}_1+\frac{3}{K}_{-1}+
4 \right].
\end{eqnarray}
For the spectral functions, we assume equal particle numbers of the two components. Then
the spectral function calculated above (and amended by the correction just described) 
refers to each component giving 

\begin{eqnarray}\label{83}
\fl \alpha _{\mbox{\small bulk}} = \frac{1}{4}\,\left[K_1+K_{-1}+\frac{1}{K}_1+
\frac{1}{K}_{-1}-4 \right],\quad\alpha _{\mbox{\small boundary}} = \frac{1}{2}\,\left[
\frac{1}{K}_1+ \frac{1}{K}_{-1}-2 \right].
\end{eqnarray}
These values agree fully with those found in \cite{FG95,MEJ97,MS00}, except for
the value of $\Delta _\perp$. Again, this results from the violation of local Lorentzian 
invariance with respect to spatial position and time in the present model.

\section{Summary}

We gave a consistent phase operator formulation for a model
of interacting one-dimensional fermions in a harmonic trap. The model is the analogue
of the Tomonaga-Luttinger model with open boundary conditions (OBC). Suitable phase operator
were identified by expressing the one-particle operators as linear odd combinations of
the basic bosonization operators in agreement with earlier results on the decomposition
of the particle density operator. The total Hamiltonian became a quadratic form
in the phase operator $\hat{\Phi}$ and the operator $\hat{\Pi}$ corresponding to the momentum 
density in spite of the presence of one-particle operators and 
the harmonic trap potential in the original fermionic Hamiltonian. These results were 
extended to the case of two components. The phase formulation bears many similarities 
to that for OBC. However, distinctive differences are also present. The spatial position
is non-linearly related to the variable in the phase fields. This spoils local Lorentzian 
invariance usually satisfied by related phase theories. In addition, form factors
appear which take account of the harmonic trap topology. Exact results for the
one-particle correlation functions were derived and used to extract bulk and boundary
critical exponents including those for the local spectral density. The values of the critical
exponents coincide with those for the OBC except for the boundary scaling 
dimension of $\hat{\psi}$.

\ack We gratefully acknowledge helpful discussions with F. Gleisberg and R. Walser 
and financial help by Deutsche Forschungsgemeinschaft.

\begin{center}
{\bf Appendix: Calculation of one-particle correlation functions}
\end{center}

\newcounter{affix}
\setcounter{equation}{0}
\setcounter{affix}{1}
\renewcommand{\theequation}{\Alph{affix}.\arabic{equation}}

\subsection{Bosonization and auxiliary correlation function}

We use the representation of auxiliary Fermi operators \cite{Schoen}

\begin{eqnarray}\label{A1}
 \hat{\psi}^+ _a (u) =\hat{O}^+(u)\, e^{-i \hat{\phi}^+ (u)}\,e^{-i \hat{\phi}
(u)}=\frac{1}{\sqrt{\eta}}\, \hat{U}^+\, e^{-i \left ( \hat{\phi}_{odd}(u)+
\hat{N}u-2 \pi \hat{\Theta}(u)  \right)}. 
\end{eqnarray}
The Klein operator $\hat{O}(u) = \exp(i \hat{N}u)\,\hat{U}$
acts on the fermion number operator $\hat{N}$. $\hat{U}$ commutes with all bosonic operators
$\left [ \hat{U}, \hat{\phi}  \right]_-= 0 = \left [ \hat{U}, \hat{\phi}^+ \right]_-$, etc. 
and changes the fermion number according to
$f( \hat{N})\, \hat{U} ^+ = \hat{U}^+ f(\hat{N}+\hat{1})$.
The contribution $\hat{N}u$ to the full phase operator in  equation (\ref{18})
leads via  equation (\ref{24}) to the ground state energy  equation (\ref{25}) and consequently to 
the chemical potential equation (\ref{33}).
We consider the one-particle correlation function of auxiliary Fermi operators

\begin{eqnarray}\label{A8}
 C_a(u,t;v) \equiv\,\, _N \langle  \hat{\psi}^+ _a (u,t) \hat{\psi}_a (v,0)\rangle _N,
\end{eqnarray}
evaluated with respect to an N-fermion state. The bosonized form is

\begin{eqnarray}\label{A9}
 C_a(u,t;v) = \frac{1}{\eta} \,\,_N \langle  \hat{O}^+(u,t)\, e^{-i (\hat{\phi}^+ (u,t)+
 \hat{\phi}(u,t) )}\, e^{i (\hat{\phi}^+ (v)+ \hat{\phi}(v))} 
\hat{O}(v) \rangle _N. 
\end{eqnarray}
The time evolution operator is governed by the separated Hamiltonian

\begin{eqnarray}\label{A10}
\hat{H} = \tilde{H} + E_0 (\hat{N})= \sum _m  m \epsilon _m\, \hat{f}^+ _m \hat{f}_m 
+ E_0 (\hat{N}),
\end{eqnarray}
made up of the ground state energy of the N-fermion system plus the Hamiltonian of
collective excitations. We thus get

\begin{eqnarray}\label{A11}
\fl C_a(u,t;v)= e^{-i(N-1)(u-v)}\,\frac{1}{\eta} \,\,
_N \langle e^{-i \left ( \hat{\phi}_{odd} (u,t)-2 \pi \hat{\Theta}(u,t)\right)}\,
\hat{U}^+ (t)\,\hat{U}\,e^{i \left (\hat{\phi}_{odd} (v)-2 \pi \hat{\Theta}(v)\right)} 
\rangle _N.
\end{eqnarray}
In order to determine $ \hat{U}^+ (t)$, we set $\hbar =1$ and
$f (\hat{N},t) \equiv \exp \left [ i E_0( \hat{N}) t \right]$ and find

\begin{eqnarray}\label{A13}
\hat{U}^+ (t)=f(\hat{N},t)\,\hat{U}^+ \,f^*( \hat{N},t) = \hat{U}^+ \,f(\hat{N}+\hat{1},t)
 \,f^*( \hat{N},t)
 \\[4mm] \nonumber
 = \hat{U}^+ \,\exp \left [  i \left ( E_0(\hat{N}+\hat{1})
-E_0 ( \hat{N}) \right)t \right]
 = \hat{U}^+ \exp \left [i \mu _{\hat{N}+\hat{1}} t \right]
 = e^{i \mu _{\hat{N}}t}\, \hat{U}^+. 
\end{eqnarray} 
Thus the time dependence of $\hat{U}^+$ is provided by the chemical potential: 
$ \hat{U}^+(t) = \exp(i \mu _{\hat{N}}t)\, \hat{U}^+$. The final result for
$C_a$ is:

\begin{eqnarray}\label{A15}
\fl C_a(u,t;v) = \frac{1}{\eta}\, e^{-i (N-1)(u-v)+ \frac{i}{\hbar}\mu _N t}\,
  \langle e^{-i \left ( \hat{\phi} _{odd}(u,t)-2 \pi  \hat{\Theta} (u,t)\right)}\,
 e^{ i \left (\hat{\phi}_{odd} (v)-2 \pi \hat{\Theta}(v)  \right)} \rangle. 
 \end{eqnarray}
The remaining expectation value is purely bosonic and can also be thermal.

\subsection{Calculation of one-particle correlation function}

The next step is to actually calculate the physical one-particle correlation function

\begin{eqnarray}\label{A16}
\fl C(z_1,t;z_2) \equiv \sum _{m,n}\psi _m (z_1) \psi _n (z_2)
\int _{-\pi}^\pi \frac{dudv}{4 \pi^2}\,  e^{imu-inv} \langle \hat{\psi}^+ _a (u,t) 
\hat{\psi}_a (v) \rangle
\\[4mm]\nonumber 
\fl = \sum _{m,n}\psi _m (z_1) \psi _n (z_2)\,\int _{-\pi}^\pi\frac{dudv}{4 \pi^2}\,  e^{imu-inv}e^{-i (N-1) 
(u-v)+i \mu _N t} \langle e^{-i \hat{\phi}^+ (u,t)} \,e^{-i \hat{\phi}(u,t)} e^{i \hat{\phi}^+
(v)} \,e^{i \hat{\phi}(v)} \rangle.  
\end{eqnarray}
This expression contains sums over harmonic oscillator wave functions $\psi _m(z)$. In 
\cite{AGW04}, a WKB method was proposed to deal with such sums: For large $N$,  
the (singular) expansion 

\begin{eqnarray}\label{A17}
\fl \sum ^\infty _{m=1} \psi _m (z_1)\,e^{imu} \rightarrow 
\left ( \frac{2 \pi ^2 \alpha^2}{N Z^2(z_1)} \right) ^{1/4} \,e^{i(N-1)u}
\\[4mm]\nonumber  
\fl \times\left \{e^{ik_F \tilde{z}(z_1)-i \pi (N-1)/2} \delta _{2 \pi} (u+u_0(z_1))
+e^{-ik_F \tilde{z}(z_1)+i \pi (N-1)/2} \delta _{2 \pi}(u-u_0(z_1)) \right\}
\end{eqnarray}
was found. Here, the following abbreviations are used

\begin{eqnarray}\label{A18}
u _0 (z) \equiv \arcsin \left(\frac{z}{L_F}\right)- \frac{\pi}{2},\quad
 \tilde{z} = \frac{1}{2} z Z(z) + \frac{1}{2}L_F  \arcsin \frac{z}{L_F}.
\end{eqnarray}
The spatial coordinates $z_1$ and $z_2$ are thus restricted to the classical region 
$(-L_F,L_F)$. The correlation function takes on the much simpler form

\begin{eqnarray}\label{A19}
\fl C(z_1,t;z_2)=\frac{\alpha \pi}{4 \pi ^2} \left ( \frac{4}{N^2 Z^2(z_1)Z^2(z_2)}
\right)^{1/4} \,e^{i \mu _N t}
\\[4mm]\nonumber
\fl \times\Bigg\{e^{ik_F (\tilde{z}(z_1)-\tilde{z}(z_2))}C_a(-u_0(z_1),t;-u_0(z_2))
+e^{-ik_F (\tilde{z}(z_1)-\tilde{z}(z_2))}C_a(u_0(z_1),t;u_0(z_2))
\\[4mm]\nonumber
\fl +e^{ik_F (\tilde{z}(z_1)+ \tilde{z}(z_2))-i \pi (N-1)}C_a (-u_0(z_1),t;u_0(z_2)) 
+e^{-ik_F(\tilde{z}(z_1)+ \tilde{z}(z_2)) + i \pi (N-1)}C_a(u_0(z_1),t;-u_0 (z_2))
\Bigg\}.
\end{eqnarray}
Note the similarity with the representation of the
one-particle correlation function for OBC in \cite{MEJ97,MS00}. However,
the argument in the operators is non-linearly related to the spatial positions 
due to the harmonic trap. In order to apply the Wick theorem 

\begin{eqnarray}\label{A19a}
\langle e^{\hat{A}}\,e^{\hat{B}} \rangle = \exp \left ( \langle \hat{A}\hat{B} \rangle 
  + \frac{1}{2} \langle \hat{A}^2 
  \rangle + \frac{1}{2} \langle \hat{B}^2  \rangle \right),  
\end{eqnarray}
the operators $\hat{A}$ and $\hat{B}$ must be homogeneous linear combinations of 
$\hat{f}$ and $\hat{f}^+$. We call the corresponding part in  equation (\ref{18}) $\hat{\Phi} _f$,
hence $\hat{\phi} _{odd}= \hat{\Phi} _f -b$, and get

\begin{eqnarray}\label{A20}
\fl C_a(u,t;v) = e^{-i(N-1)(u-v)+ i \mu _N t+ib(u)-ib(v)}\, \langle e^{-i \hat{\Phi}_f(u,t) 
+ 2 \pi i \hat{\Theta}(u,t)}\,e^{i \hat{\Phi}_f(v)-2 \pi i \hat{\Theta}(v)} \rangle.
\end{eqnarray}
Thus 
$\hat{A} \equiv - i \hat{\Phi}_f(u,t)+2 \pi i \hat{\Theta}(u,t) = \hat{A}(u,t),\quad
\hat{B} \equiv i \hat{\Phi}_f (v)- 2 \pi i \hat{\Theta}(v)= - \hat{A}(v,0)$.
 
We begin with the case of zero temperature when only $\langle \hat{f}_m \hat{f}^+ _n \rangle _0 
=\delta _{m,n} $ survives. Then simple but lengthy calculations using  equation (\ref{15c}) and
$1/K_m \rightarrow 1+(1-K)\,\exp(-mr)/K$ and the coupling constants

\begin{eqnarray}\label{A23}
\alpha _0 = \frac{1}{4} \left[ \frac{1}{K}-K \right],\quad\gamma _0=\frac{(1-K)^2}{4 K},
\end{eqnarray}
give

\begin{eqnarray}\label{A24}
\exp\left(\frac{1}{2}\langle \hat{A}^2 (u)\rangle _0 \right) = \sqrt{\eta}\, r^{\gamma _0} 
\left \{{\cal{D}}(2u + i r) {\cal{D}}(-2u +i r )    \right \} ^{-\alpha _0/2},
\end{eqnarray}
with ${\cal{D}} (s)\equiv 1/(1-\exp(is))$.
The correlation function $\langle \hat{A} (u,t) \hat{B} (v)\rangle _0 = - \langle \hat{A}(u,t)
\hat{A} (v,0) \rangle _0$ turns out to be
 
 \begin{eqnarray}\label{A26}
 \langle \hat{A} (u,t)\hat{B}(v)\rangle _0  = \ln {\cal{D}} (u-v-\epsilon t + i \eta)
 \\[4mm]\nonumber
 + \gamma _0 \left \{ \ln {\cal{D}} (u-v-\epsilon t - i r) + \ln {\cal{D}}(v-u-
 \epsilon t + i r)  \right \}
 \\[4mm]\nonumber
 + \alpha _0 [ \ln {\cal{D}}(u+v-\epsilon t + i r) + \ln {\cal{D}} (-u-v-\epsilon t +
 ir)].
    \end{eqnarray}
The final zero temperature result is
 
 \begin{eqnarray}\label{A28}
\fl C_a(u,t;v)=\langle \hat{\psi}^+ _a (u,t)\hat{\psi}_a (v)\rangle _0 = e^{-i(N-1)(u-v)
+i \mu _N t+ib(u)-ib(v)} \,r ^{2 \gamma _0} 
\\[4mm]\nonumber
\fl   \times\frac{ \left[{\cal{D}}(u-v-\epsilon t + ir)\,{\cal{D}}(v-u-\epsilon t + ir) \right]
 ^{\gamma _0}}{1-e^{-\eta + i (u-v-\epsilon t)}} 
\left\{\frac{ {\cal{D}}(u+v-\epsilon t + i r)\,{\cal{D}}(-u-v-\epsilon t+i r)}
{|{\cal{D}}(2u+ir){\cal{D}}(2v+ir)|}\right\}^{\alpha _0}.  
\end{eqnarray}
 Note that in contrast to \cite{MEJ97} two small parameters appear in  equation (\ref{A28}),
 the positive infinitesimal $\eta$ and $r$ which is finite in the finite system. In 
 addition, the one-particle operators in the Hamiltonian lead to additional phase 
 factors which can be expressed as:
$e^{i(b(u)-b(v))}= \left \{{\cal{D}}(2u+ir)\,{\cal{D}}(-2v+ir)/({\cal{D}}(-2u+ir)\,
{\cal{D}}(2v+ir))   \right\}^{\kappa _0}$.

The main difference of  equation (\ref{A28}) to the formula (25) in \cite{MEJ97} 
is the appearance of the auxiliary variable
$u$ (or $u_0(z))$ in place of the position. Otherwise, our result (\ref{A28}) is in
direct correspondence to that formula. As a consequence, the 
finite temperature calculation of \cite{MEJ97} can be adopted to obtain the result for the 
canonical ensemble by making the substitution
 
 \begin{eqnarray}\label{A29}
 {\cal{D}}(s)=\frac{1}{1-e^{is}}\rightarrow
\tilde{\cal{D}}(s)={\cal{D}}(s)/\Pi _{k=1}\left[1+\left(\frac{\sin(s/2)}
    {\sinh(k \beta \epsilon/2)}\right)^2 \right],
\end{eqnarray}
with $\beta^{-1}=k_B T$. We will not pursue the finite temperature case.
In the same way, we find
 
 \begin{eqnarray}\label{A30}
\fl \langle \hat{\psi} _a (u,t)\hat{\psi}^+_a (v)\rangle _0 = e^{iN (u-v)-i \mu _{N+1} t
-ib(u)+ib(v)} \,r ^{2 \gamma _0} 
\\[4mm]\nonumber
\fl \times\frac{ \left[{\cal{D}}(u-v-\epsilon t + ir)\,{\cal{D}}(v-u-\epsilon t + ir) \right]
 ^{\gamma _0}}{1-e^{-\eta + i (u-v-\epsilon t)}} 
 \left\{ \frac{{\cal{D}}(u+v-\epsilon t + i r)\,{\cal{D}}(-u-v-\epsilon t+i r)}
{|{\cal{D}}(2u+ir){\cal{D}}(2v+ir)|}\right\} ^{\alpha _0}. 
 \end{eqnarray}

% Unpublished appendix B
\begin{center}
{\bf Appendix B: Backscattering between the two components}
\end{center}

\setcounter{equation}{0}
\setcounter{affix}{2}
\renewcommand{\theequation}{\Alph{affix}.\arabic{equation}}

In the approximation $V_{c\perp}(|m|)\rightarrow V_{c\perp}$, the associated two-particle operator
in second quantization is
\begin{eqnarray}\label{B1}
\hat{V}_{\perp}=\frac{1}{2}V_{c\perp} \sum_{\sigma}\int^{\pi}_{-\pi} \frac{dv}{2\pi}
\hat{\psi}^{+}_{a -\sigma}(v)\hat{\psi}^{+}_{a \sigma}(-v)\hat{\psi} _{a \sigma} (v)\hat{\psi} _{a -\sigma} (-v).
\end{eqnarray}
Using
\begin{eqnarray}\label{B2}
\hat{\psi}^{+}_{a \sigma}(v)=e^{-i(\hat{N}_\sigma-1)v}\,\hat{U}^{+}_\sigma\frac{1}{\sqrt{\eta}}
e^{-i \hat{\phi}^+_\sigma (v)-i \hat{\phi}_\sigma (v)}
\end{eqnarray}
together with equation (\ref{6}), and utilizing $[\hat{U}_\sigma,\hat{N}_{\sigma^{\prime}}]
=\delta_{\sigma,\sigma^{\prime}}\hat{U}_\sigma$, we obtain the following phase representation of 
$\hat{V}_{\perp}$
\begin{eqnarray}\label{B3}
\fl \hat{V}_{\perp}=\frac{V_{c\perp}}{4\pi\eta^2} \sum_{\sigma}\int^{\pi}_{-\pi} d u
\exp[2iu(\hat{N}_{\sigma}-\hat{N}_{-\sigma})]
\exp[2i(\hat{\varphi}_{\sigma,odd}(u)-\hat{\varphi}_{-\sigma,odd}(u))],
\end{eqnarray}
with new phase operators,
\begin{eqnarray}\label{B4}
 \hat{\varphi}_{\sigma,odd}(u) =\sum_{\nu}\sum^{\infty}_{m=1}
\sigma^{\frac{1-\nu}{2}} \sqrt{\frac{e^{-m \eta}}
{2m}}\,\sin(m u)\,\left(\hat{d}_{m \nu} +\hat{d}^+ _{m \nu} \right).
\end{eqnarray}

Because of $\hat{\varphi}_{\sigma,odd}(u)-\hat{\varphi}_{-\sigma,odd}(u)\equiv \sigma \sqrt{2}
(\hat{\Phi}_{-1}-u\hat{N}_{\nu=-1} )$ the simpler form
\begin{eqnarray}\label{B5}
\hat{V}_{\perp}=\frac{V_{c\perp}}{2\pi\eta^2} \int^{\pi}_{-\pi} d u \cos\left[\sqrt{8}\hat{\Phi}_{-1}(u)\right]
\end{eqnarray}
is obtained. As expected, only the composition part is affected by backscattering between components.

Finally, it is noted that that normal ordered form of equation (\ref{B5}) is
\begin{eqnarray}\label{B6}
\hat{V}_{\perp}=V_{c\perp} \int^{\pi}_{-\pi} \frac{du}{2\pi}
:\frac{\cos\left[\sqrt{8}\hat{\Phi}_{-1}(u)\right]}{1+e^{-2\eta}-e^{-\eta}\cos(2u)}:.
\end{eqnarray}

\end{document}